\documentclass[prl,aps,twocolumn,superscriptaddress,notitlepage]{revtex4-1}
\usepackage{amssymb,amsfonts,amsmath,amstext,graphicx,hyperref}
\usepackage{tikz}
\usepackage{graphicx}
\usepackage[normalem]{ulem}
\hypersetup{
	colorlinks=true,
	linkcolor=blue,
	filecolor=magenta,      
	urlcolor=cyan,
}
\usepackage{xcolor}

\usepackage[normalem]{ulem}
\newcommand{\cut}[1]{}

\usepackage{physics}
\def\braket#1{\langle{#1}\rangle}
\def\lket#1{\vert#1\rangle\hspace{-1mm}\rangle}
\def\lbra#1{\langle\hspace{-1mm}\langle#1\vert}

\begin{document}

\title{Measurement Induced Continuous Time Crystals}

\author{Midhun Krishna}
\email{midhunkrishna@iitb.ac.in}
\affiliation{Department of Physics, Indian Institute of Technology-Bombay, Powai, Mumbai 400076, India}

\author{Parvinder Solanki}
\email{psolanki@phy.iitb.ac.in}
\affiliation{Department of Physics, Indian Institute of Technology-Bombay, Powai, Mumbai 400076, India}

\author{Michal Hajdu\v{s}ek}
\email{michal@sfc.wide.ad.jp}
\affiliation{Keio University Shonan Fujisawa Campus, 5322 Endo, Fujisawa, Kanagawa 252-0882, Japan}
\affiliation{Keio University Quantum Computing Center, 3-14-1 Hiyoshi, Kohoku, Yokohama, Kanagawa 223-8522, Japan}

\author{Sai Vinjanampathy}
\email{sai@phy.iitb.ac.in}
\affiliation{Department of Physics, Indian Institute of Technology-Bombay, Powai, Mumbai 400076, India}
\affiliation{Centre for Quantum Technologies, National University of Singapore, 3 Science Drive 2, 117543 Singapore, Singapore}

\date{\today}

\begin{abstract}
Strong measurements usually restrict the dynamics of measured finite dimensional systems to the Zeno subspace, where subsequent evolution is unitary due to the suppression of dissipative terms.
Here we show  qualitatively different behaviour due to the competition between strong measurements and the thermodynamic limit, inducing a time-translation symmetry breaking phase transition resulting in a continuous time crystal.
We consider a spin star model, where the central spin is subject to a strong continuous measurement, and qualify the dynamic behaviour of the system in various parameter regimes.
We show that above a critical value of measurement strength, the magnetization of the thermodynamically large ancilla spins develops limit cycle oscillations.
Our result also demonstrates that a coherent drive is not necessary in order to induce continuous time-translation symmetry breaking.

\end{abstract} 

\maketitle

\textit{Introduction.---}
Quantum measurements are a central aspect of open quantum system dynamics with applications to quantum computing \cite{raussendorf2001oneway}, sensing \cite{degen2017quantum}, communications \cite{khatri2020principles} and cryptography \cite{gisin2002quantum}.  Since measurements are non-unitary and act on a timescale different than Hamiltonian evolution, novel effects such as weak-value amplification \cite{aharonov1988result}
, quantum Zeno \cite{misra1977zeno} and anti-Zeno \cite{antizeno,mukherjee2020anti} effects are observed in finite systems. The Zeno freezing of the dynamics does not preclude subsequent dynamical behaviour of the system, since a quantum system that is strongly measured can still evolve unitarily in its Zeno subspace \cite{Burgarth2019generalized,Zanardi,facchi2002quantum,facchi2008quantum}.
The situation becomes more subtle for many-body systems, where competition between measurement strength and thermodynamic limit can create novel phases. While such a competition has been studied to understand emergent steady states of quantum systems either measured or coupled to dissipative baths which model measurements \cite{kessler2012dissipative,secli2022steady}, it is an open question to know if time independent measurements \textit{alone} can induce continuous time translation symmetry breaking in open quantum systems in the thermodynamic limit. In this manuscript, we answer this question in the affirmative by inducing a continuous time crystal entirely by measurements. 

Time crystals are novel phases of matter with broken time-translation symmetry \cite{sacha2017time,else2020discrete}. The earliest examples of these many-body non-equilibrium systems displayed breaking of the \emph{discrete} time-translation symmetry in closed and open systems \cite{else2016floquet,russomano2017floquet,surace2019floquet,khasseh2019many,gong2018discrete,lazarides2020time,RieraCampeny2020time,cabot2022metastable,tuquero2022dissipative} and have been observed experimentally in a number of physical platforms \cite{zhang2017observation,choi2017observation,pal2018temporal,rovny2018observation,smits2018observation,kyprianidis2021observation,taheri2022time,kessler2021observation,mi2022time,frey2021realization,randall2021manybody}.
It has been recently realized that dissipation plays a crucial role in breaking of the continuous time-translation symmetry in otherwise quiescent systems, leading to \emph{continuous time crystals} (CTCs) \cite{iemini2018boundary,tucker2018shattered,buca2019nonstationary,zhu2019dicke,lledo2019driven,seibold2020dissipative,prazeres2021boundary,piccitto2021symmetries,carollo2022exact,hajdusek2022seeding,nakanishi2022dissipative}. Here, the continuous symmetry is broken in the thermodynamic limit witnessed by the emergence of a self-organized oscillating steady state. These limit cycle oscillations were recently observed in a continuously pumped dissipative atom-cavity system \cite{kongkhambut2022observation}. While discrete time crystals are induced by periodic driving, all existing models of continuous time crystals are induced by the presence of a constant external drive.
In this manuscript, we also demonstrate that the drive can be elimininated and replaced entirely by a time-translation invariant measurement scheme which provides both the drive and dissipators necessary in the thermodynamic limit to induce a dissipative phase transition.

\textit{Zeno-Induced Dissipative Phase Transitions.---}
Physical systems that exhibit dissipative phase transitions involve competition between coherent and dissipative terms.
It is however not always the case that these terms scale extensively with the size of the system.
Typical approach is to rescale the non-extensive terms \cite{defenu2021long,anteneodo1998breakdown,kirton2019introduction}, which amounts to demanding that the dissipation rates scale inversely with the system size.
Examples of this strategy can be found in both closed and open collective models \cite{bastidas2012nonequilibrium,ferreira2019lipkin,hwang2016quantum,garbe2017superradiant,bakker2022driven,minganti2018spectral}, with CTCs being no exception.

Typical derivations of Markovian master equations from microscopic physics do not provide a path to this rescaling since system environment interactions are rarely tunable.  In contrast to this, variation of measurement strength with the size of the system as a power-law is more readily understood as a feasible control strategy \cite{garbe2017superradiant}.
This opens the attractive possibility of using measurements as a natural path to arbitrary power-law variation of the effective dissipation strength of open quantum systems.
This, alongside typically controllable Hamiltonian terms provides complete control of the dissipative phase transition.

Consider a subsystem of a thermodynamically large system being subjected to strong measurements.
As detailed below, it can be shown that the measurement induces a Lindblad evolution on the remainder of the system \cite{popkov18,kessler2012generalized,kessler2012dissipative,popkov21}.
The strength of the Lindblad evolution grows weaker with  measurement strength.
This implies that if the measurement strength scales suitably with the size of the unmeasured system, it is possible for extensive Hamiltonian terms to compete with the effective dissipation, causing qualitatively large changes in the steady state properties in the thermodynamic limit, heralding a dissipative phase transition.

\textit{Spin Star Model.---}
\begin{figure}
    \centering
    \includegraphics[width=0.9\columnwidth]{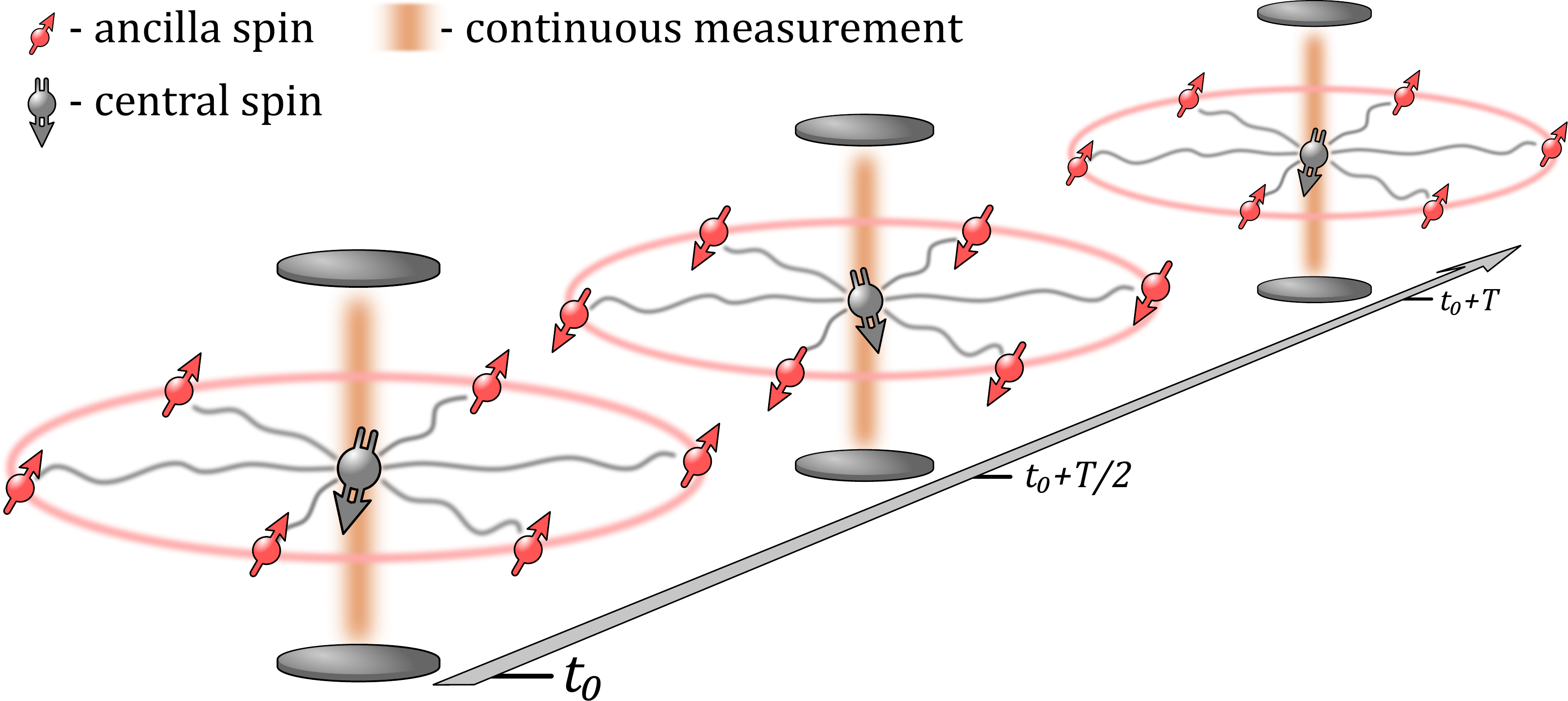}
    \caption{Spin star model, where a single central spin interacts with multiple ancilla spins. Strong continuous measurement of the central spin induces transient oscillations of magnetization for finite number of ancilla spins, and leads to time-translation symmetry broken phase in the thermodynamic limit. Period of these oscillations is given by $T=2\pi/\Omega_z$, and is explicitly calculated below.}
    \label{fig:figure1_CSM}
\end{figure}
In order to demonstrate how measurements can induce the time-crystalline phase, we henceforth focus on a specific model given by the spin star system \cite{hutton2004mediated,barnes2012nonperturbative,bortz2007exact}, as shown in Fig.~\ref{fig:figure1_CSM}.
A central spin interacting with $N$ ancilla spins is described by the following Hamiltonian,
\begin{equation}\label{eq:CSM}
    \hat{H}_{T} = \omega_{c} \hat{S}_{z} + \omega_{a} \hat{I}_{z} + \vec{S}\cdot\mathbf{J}\cdot\vec{I} ,
\end{equation}
where $\hat{I}_{\alpha}= \sum_{k}\hat{\sigma}^{k}_{\alpha}/2$ denotes the collective spin operator of the ancilla spins, $\vec{I} = ( \hat{I}_{x},\hat{I}_{y},\hat{I}_{z})^{\intercal}$, $\hat{S}_{\alpha} = \hat{\sigma}_{\alpha}/2$ denotes the spin operator of the central spin, and $\vec{S} = ( \hat{S}_{x},\hat{S}_{y},\hat{S}_{z})$.
The natural frequencies of the central spin and collective ancilla spins are $\omega_{c} \text{ and } \omega_{a}$, respectively.
The interaction between the central spin and the ancilla spins is characterized by the coupling matrix $\mathbf{J}$.
The central spin is coupled to a measurement apparatus modeled as a zero temperature bath, inducing an open system evolution \cite{breuer2002theory}
of the spin star system given by 
\begin{equation}\label{eq:LME}
    \dot{\zeta} = -i[\hat{H}_{T},\zeta] +\Gamma\mathcal{D}[\hat{S}^{-}]\zeta.
\end{equation}
Here $\zeta$ is the state of the spin star system and $\mathcal{D}[\hat{O}]\zeta = \hat{O}\zeta \hat{O}^{\dagger}- \{\hat{O}^{\dagger}\hat{O},\zeta\}/2$ is a standard Lindblad dissipator that models the effect of the measurement acting on the central spin with strength $\Gamma$.

We consider the dissipative spin star system in the Zeno limit of strong measurement and treat the Hamiltonian part perturbatively \cite{popkov18,kessler2012generalized} to arrive at effective dynamics for the ancilla spins.
Keeping leading order terms produces the effective master equation for ancilla spins in Lindblad form (see appendix~A),
\begin{equation}\label{eq:EME}
    \dot{\rho} = -i[\hat{H},\rho] + \frac{1}{\Gamma}\mathcal{D}[\hat{L}]\rho +\mathcal{O}\left(\frac{1}{\Gamma^{2}}\right).
\end{equation}
Here, $\rho$ is the reduced state of the ancilla spins after tracing out the central spin, $\hat{H} = \omega_{a} \hat{I}_{z} - \frac{1}{2}\sum_{\alpha} J_{z\alpha} \hat{I}_{\alpha} $ is the effective Hamiltonian, and $\hat{L} = \sum_{\alpha,\beta}(J_{x\alpha}\hat{I}_{\alpha} -i J_{y\beta}\hat{I}_{\beta}) $ is the jump operator for the effective dissipation. The $\mathcal{O}(1/\Gamma^2)$ correction in Eq.~(\ref{eq:EME}) can be neglected when $\| \hat{H}_{T} \|_{\infty} \ll \Gamma$, where $\|.\|_{\infty}$ is the infinity norm (see appendix A).
We now demonstrate this by studying the Liouville eigenspectrum of the full spin-star system alongside the ancilla subsystem.

\textit{Liouville Eigenspectra \& Zeno Limit.---}
The Liouville eigenspectrum encodes both the transient and steady state behaviour of open system dynamics \cite{albert2014symmetries,RieraCampeny2020time,solanki2022role,jaseem2020generalized}.
Both master equations for the open spin star system in Eq.~(\ref{eq:LME}) and the ancilla spins in Eq.~(\ref{eq:EME}) can be written in the form of vectorized density matrices as $\lket{\dot{\rho}} = \mathcal{L}\lket{\rho}$ \cite{albert2014symmetries}. The spectral decomposition of a Liouvillian can be expressed as $\mathcal{L} = \sum_{k} \lambda_{k} \lket{l_{k}}\lbra{r_{k}}$, where $\lambda_{k} = \alpha_{k} +i\beta_{k}$  are the complex eigenvalues with corresponding left and right eigenvectors $\lket{l_{k}}$ and $\lket{r_{k}}$. 

The dissipator $\mathcal{D}[\hat{S}^{-}]$ when restricted to the central spin has four eigenstates with eigenvalues $\{0,-1/2,-1/2,-1\}$.
In the full Liouville space of the spin star system, when the Hamiltonian part is not considered, these eigenvalues become $\{0,-\Gamma/2,-\Gamma/2,-\Gamma\}$, each having degeneracy given by the dimension of the ancilla Liouville space.
The Hamiltonian in Eq.~(\ref{eq:LME}) does not commute with the dissipative term, therefore lifting the degeneracy of the full Liouvillian.
Hence the Liouville eigenspectrum of the entire system shows the eigenvalues occupy vertical stripes with horizontal separation of $\mathcal{O}(\Gamma)$ between them, and a spread within each stripe of $\mathcal{O}(1/\Gamma)$, as seen in Fig.~\ref{fig:Finite_Zeno}. Since the eigenvalues of $\mathcal{D}[\hat{S}^-]$ are purely real, the vertical spread of the spectrum depends only on the parameters of the Hamiltonian and is independent of the measurement strength $\Gamma$. We label the eigenvalues of the full Liouvillian of the spin star system as $\{\lambda_k^{(\mu)}\}_{\mu = \{0,1,2,3\}}$, with superscript $\mu$ denoting the vertical stripe of the eigenvalue and subscript $k < 4(N+1)^2$ enumerating the eigenvalues of the full Liouvillian within the symmetric space of the ancilla spins.

\begin{figure*}[htp!]
    \centering
    \includegraphics[width=1\linewidth]{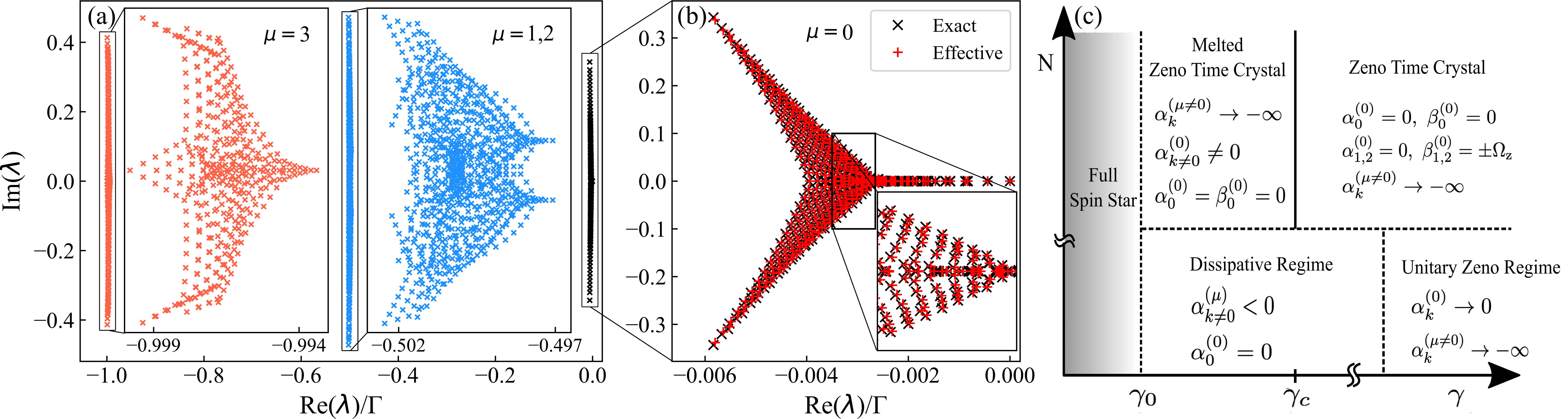}
    \caption{Eigenspectrum of the spin star Liouvillian superoperator has four stripes ($\mu =0,1,2,3$) corresponding to different eigenvalues of Lindblad dissipator as shown in (a). Subfigure (b) shows that stripe $\mu = 0$ of the spin star Liouvillian ($\times$ in black) matches well with the effective ancilla Liouvillian ($+$ in red ) for moderate reduced measurement strength of $\gamma = 15$. Corresponding non-zero Hamiltonian parameters are $\omega_{c}/J_{xx} = 0.1$, $\omega_{a}/J_{xx} = 0.01$ $J_{yy}/J_{xx} = 1$, $J_{zx}/J_{xx} = 0.01$ and $N=20$. Competition between reduced measurement strength $\gamma$ and system size $N$ gives rise to regions of qualitatively different behaviour in the parameter space as presented in (c). The dotted lines are soft boundaries with the kinks on $\gamma$ and $N$ axis representing the Zeno limit and thermodynamic limit respectively. The effective ancilla master equation is valid in the region on right side of  vertical dotted line at $\gamma_{0}$. The presence of phase transition is represented by the solid vertical like at critical point $\gamma_{c}$.}
    \label{fig:Finite_Zeno}
\end{figure*}

In the Zeno limit of strong measurement, the dynamics is confined in the steady state subspace of the dissipator, and is usually referred to as the Zeno subspace \cite{Burgarth2019generalized,facchi2002quantum,facchi2008quantum}.
In the case of Eq.~(\ref{eq:LME}), the Zeno subspace is given by the central spin being in the ground state.
This corresponds to the steady state stripe of the eigenspectrum with $\mu=0$.
We note that the effective dynamics inside the Zeno subspace described by Eq.~(\ref{eq:EME}) has the Lindblad structure with weak dissipation.
The difference in the eigenvalues of the full spin-star Liouvillian and the ancilla Liouvillian scales as $\Gamma^{-2}$, implying nearly overlapping spectra for large enough measurement strengths \cite{popkov21,popkov18}. To compare the effect of the measurement strength and the system size on the dynamics, we define reduced measurement strength $\gamma$, such that $\Gamma = \gamma I$, with $I=N/2$ being the total spin of the ancilla spins. 

In order to neglect the higher-order term in Eq.~(\ref{eq:EME}), and arrive at an effective Lindblad evolution for the ancilla spins, we require the reduced measurement strength $\gamma$ to be much greater than a threshold strength $\gamma_{0} \approx \max\{J_{ij}, \omega_{c},\omega_{a}\}$.
For the parameters of the Hamiltonian chosen in Fig.~\ref{fig:Finite_Zeno}(a), moderate value of  $\gamma=15$ shows segregation of the eigenspectrum of spin star Liouvillian into distinct stripes.
Fig.~\ref{fig:Finite_Zeno}(b) shows excellent agreement between the Liouville eigenspectrum of the full and effective evolutions given by Eq.~(\ref{eq:LME}) and Eq.~(\ref{eq:EME}), respectively.

The Zeno limit of $\Gamma \rightarrow \infty$ can be achieved either by keeping the system size fixed and increasing the reduced measurement strength ($\gamma \rightarrow \infty$) or by keeping the reduced measurement strength constant and going to the thermodynamic limit ($N \rightarrow \infty$).
Depending on the value of $\gamma$ and $N$, the spin star model shows distinct dynamical behaviour as sketched in Fig.~\ref{fig:Finite_Zeno}(c).
We explore these further below.
Subsequent discussions are made in the region where $\gamma \gg \gamma_{0}$, where the ancilla master equation is valid.


\textit{Finite spin-star system.---} As the strength of measurements on the central spin increases, the effective dissipation on the collective ancilla spin becomes weaker.
For a finite ancilla size depending on the value of $\gamma$, we observe two distinct regimes.
These regimes are referred to as \emph{dissipative regime} and \emph{unitary Zeno regime} in Fig.~\ref{fig:Finite_Zeno}(c).
In the dissipative regime, all the eigenvalues of the spin star Liouvillian, except the one corresponding to a steady state, have finite real part, $\alpha_{k\neq0}^{\mu} \neq 0$.
Hence the dynamics takes arbitrary initial symmetric states of the ancilla spins to the steady state characterized by $\lambda_{0} = 0$. 
In contrast, in the unitary Zeno regime when $\gamma \rightarrow \infty$, the dissipation inside the Zeno subspace vanishes and the dynamics of collective ancilla spins is unitary \cite{Zanardi,Burgarth2019generalized}.
In terms of the Liouville eigenspectum of the spin star system, the unitary Zeno regime can be viewed as the collapse of steady state stripe onto the imaginary axis, $\alpha_{k}^{0} \rightarrow 1/\Gamma \rightarrow 0$, and the divergence of the eigenvalues from the other stripes towards negative infinity $\alpha_{k}^{\mu \neq 0} \rightarrow -\Gamma \rightarrow -\infty$.
In this limit, the measurement overwhelms the dynamics of the central spin and drives it to a steady state given by $\braket{\vec{S}} = (0,0,-1/2)$, essentially decoupling it from the ancilla spins.
We now show that variation of measurement strength induces strikingly different behaviour in the thermodynamic limit.

\textit{Thermodynamic Limit.---}
In contrast to the previous section, where we approached the Zeno limit by $\gamma\rightarrow\infty$ for a finite $N$, we choose a fixed value of the reduced measurement strength $\gamma\gg\gamma_0$ and let the number of the ancilla spins diverge, $N\rightarrow\infty$.
Doing this opens the possibility for a dissipative phase transition to occur, as both the unitary and dissipative parts of the master equation~(\ref{eq:EME}) scale extensively with the size of the system. To illustrate this explicitly, we make a particular choice for the coupling matrix $\mathbf{J}$ in the Hamiltonian with non-zero entries $J_{xx} = J_{yy}, J_{zz} = 2\omega_{a} \text{ and } J_{zx} = -2\Omega$. Such anisotropic terms can be engineered by directional hopping in Fermionic models \cite{hassanRMP,hassan,anisotropic} or optical lattices.
 This choice of parameters reduces the ancilla master equation~(\ref{eq:EME}) to the following form,
\begin{equation}\label{eq:ZTC}
    \dot{\rho} = -i\Omega[\hat{I}_{x},\rho] +\frac{\kappa}{I} \mathcal{D}[\hat{I}_{-}]\rho, 
\end{equation}
where $\kappa := J_{xx}^2/\gamma$ is the effective dissipation rate. This is the well known driven Dicke model of continuous time crystallization \cite{iemini2018boundary,prazeres2021boundary,piccitto2021symmetries,carollo2022exact}. 
The driven Dicke master equation~(\ref{eq:ZTC}) displays a dissipative phase transition between a stationary melted phase for $\Omega/\kappa \leq 1$ and a continuous time crystal phase for $\Omega/\kappa > 1$ \cite{iemini2018boundary}, where the system self-organizes into a steady state with persistent limit-cycle oscillations.

We note that the phase transition occurs at the critical reduced measurement strength $\gamma_{c} = J_{xx}^{2} /\Omega$. This implies that for reduced measurement strength $\gamma_{0} \ll \gamma \leq \gamma_{c}$, the ancilla spins are in a stationary melted phase, where the effective dissipation dominates over the coherent evolution.
In contrast,  when $\gamma > \gamma_{c} \gg \gamma_0$, the ancilla spins enter a continuous time crystal phase with persistent oscillations of their magnetization, breaking the time-translation symmetry of the underlying dynamics, as seen in Fig.~\ref{fig:Finite_Zeno}(c).

\begin{figure}
    \centering
    \includegraphics[width=\linewidth]{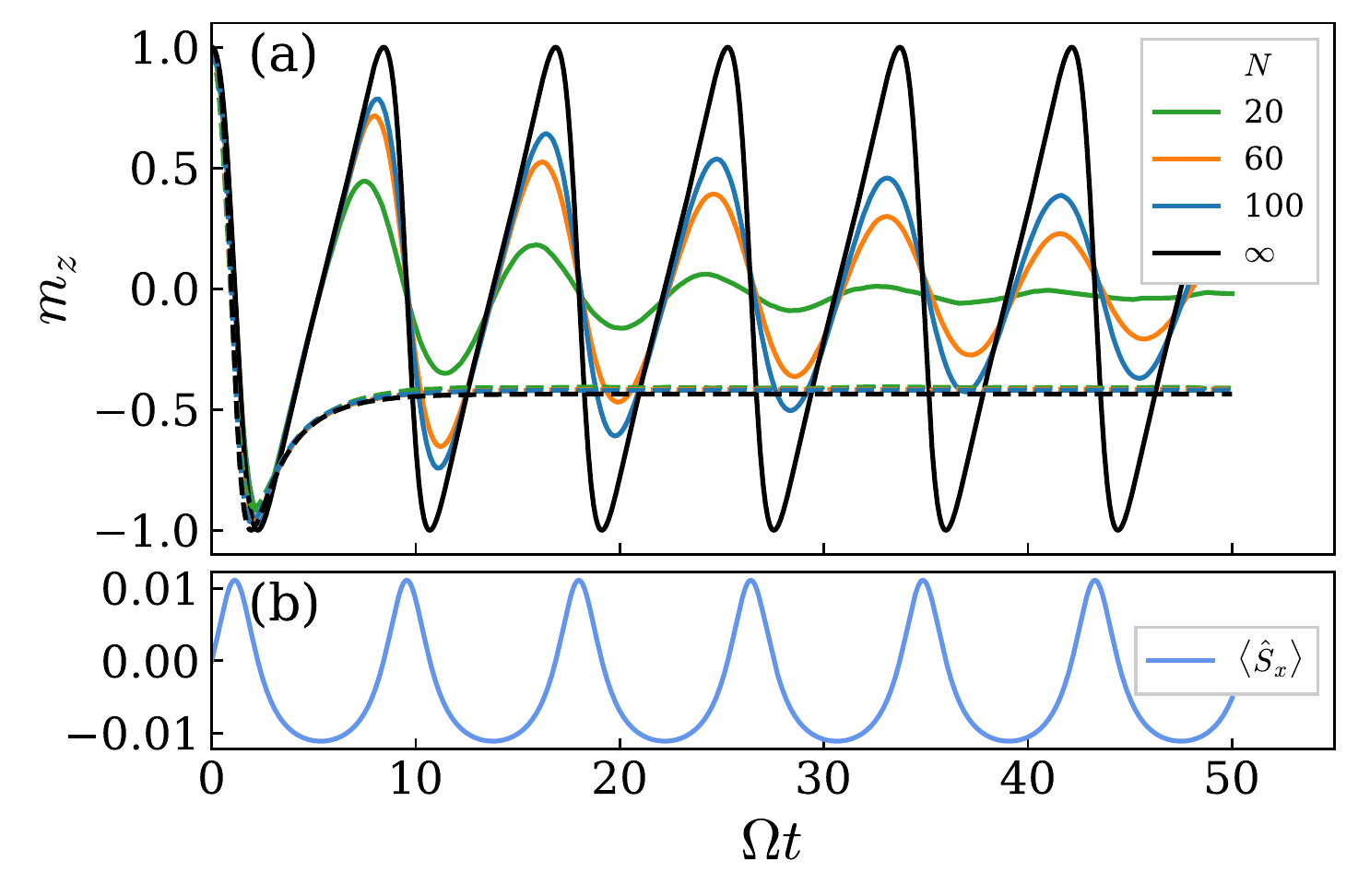}
    \caption{(a) Time evolution of rescaled ancilla spin operator $m_{z}$ for varying numbers of ancilla spins for initial state $\vec{m}=(0,0,1)$. The solid lines correspond to the dynamics in the ZTC phase with $\Omega/\kappa = 1.5$, and show longer-lived oscillations as N increases. These oscillations become persistent in the thermodynamic limit. The dashed lines correspond to the melted phase with $\Omega/\kappa = 0.9$, for which $m_{z}$ settles down to a saddle fixed point. (b) Magnetization $\langle \hat{S}_x \rangle$ of the central spin oscillates with the same frequency as the ancilla spins.}
    \label{fig:ZTC_Oscillations}
\end{figure}

The emergence of limit cycles in continuous time crystals is traditionally understood as a competition between oscillations-inducing coherent drive and oscillations-destroying dissipation \cite{iemini2018boundary,buca2019nonstationary,kessler2019emergent,prazeres2021boundary}.
We note the absence of any coherent drive in the full spin star model of Eq.~(\ref{eq:LME}).
The key ingredients in our model are anisotropic coupling of the central spin to the ancilla spins given by finite off-diagonal element of the coupling matrix $\mathbf{J}$ and the Zeno limit given by diverging measurement strength $\Gamma$.
In order to distinguish the origin of the effective model in Eq.~(\ref{eq:ZTC}) from more traditional models in literature, we refer to this measurement-induced symmetry-broken phase as a \emph{Zeno time crystal} (ZTC).

The ZTC phase can be witnessed by examining the Liouville eigenspectrum of the full spin star model in Eq.~(\ref{eq:LME}).
In the Zeno limit taken by $N\rightarrow\infty$ for a fixed reduced measurement strength $\gamma$, the asymptotic dynamics settles into oscillating coherences.
These are eigenstates of Eq.~(\ref{eq:LME}) located within $\mu=0$ stripe with pure imaginary eigenvalues $\lambda_{1,2} = \pm i \Omega_Z$, where $\Omega_Z = \sqrt{(J_{xx}^{2}/\gamma)^2 - \Omega^{2}}$ is the frequency of the limit-cycle oscillations.
Eigenstates with eigenvalues located within the remaining stripes, $\mu\neq0$, have eigenvalues with non-vanishing real parts and therefore have no effect on late-time dynamics.

We plot the dynamics of the rescaled magnetization of the ancilla spins, $m_z \equiv \langle\hat{I}_z\rangle / N$, in Fig.~\ref{fig:ZTC_Oscillations}(a) for increasing values of $N$.
We observe that for rescaled measurement strength $\gamma \leq \gamma_c$, $m_z$ settles into a steady value.
This is represented by the dashed lines and corresponds to the melted ZTC phase of Fig.~\ref{fig:Finite_Zeno}(c).
On the other hand, for $\gamma > \gamma_c$, the rescaled magnetization of the ancilla spins displays transient oscillations for finite $N$ as represented by the solid lines in Fig.~\ref{fig:ZTC_Oscillations}(a).
The stability of these oscillations increases with the number of ancilla spins $N$.
In the thermodynamic limit, $N\rightarrow\infty$, these oscillations become persistent, signaling a genuine phase with broken time-translation symmetry.
The full system of nonlinear mean-field equations for $\langle\hat{I}_{\alpha}\rangle$ as well as the central spin $\langle\hat{S}_{\alpha}\rangle$ can be found in the appendix~B.

For completeness, we examine the dynamics of the central spin in the Zeno limit and diverging number of ancilla spins $N$.
The evolution of the central spin is dominated by the measurement strength $\Gamma$, resulting in rapid damping of $\langle\vec{S}\rangle$.
Solving the mean field dynamics (see appendix~B) of the coupled system in the Zeno limit using adiabatic elimination \cite{haken1977synergetics} leads to the steady-state solution for the central spin given by $\langle\vec{S}\rangle \approx ( -J_{xx}m_y/\gamma, J_{xx}m_x/\gamma, -1/2 )$.
Since the rescaled magnetization of the ancilla spins, $m_x$ and $m_y$, display limit-cycle oscillations in the ZTC phase, so does the magnetization of the central spin, as shown in Fig.~\ref{fig:ZTC_Oscillations}(b) where we plot the steady-state dynamics of $\langle\hat{S}_x\rangle$.
This demonstrates that time-translation symmetry breaking does not only occur in the ancilla spins but in the central spin as well.
We see that provided $\gamma > \gamma_c$, the central spin oscillates with the same frequency as the ancilla spins.

\textit{Conclusions.---}
In this manuscript, we provide a path to observe continuous time-translation symmetry breaking in many-body phase transitions induced by static quantum measurements alone. We demonstrate how to observe Zeno time crystals in spin star system, a model well studied both theoretically \cite{fischer2009hyperfine,barnes2012nonperturbative,faribault2013integrability,van2014competing,bortz2007exact,seifert2016persisting} and experimentally \cite{mahesh2021star,suter2017single,bradley2019ten,poggiali2017measurement}, forming a suitable platform for stable quantum technologies. The relationship between phase coherence, limit cycles and quantum technologies has been studied recently in a number of physical systems theoretically and experimentally  \cite{sonar2018squeezing,jaseem2020quantum,laskar2020observation,jaseem2020generalized,krithika2022observation}. Our result is counter-intuitive in the role quantum measurements have played in phase transitions.
Unlike previous models which involve explicit coherent drive, we can observe time-crystallinity directly from strong measurement of a subsystem, which provides both the coherent and dissipative terms for the time crystal.
Furthermore, these local measurements are usually thought of as localizing the wavefunction and hence disrupting the otherwise entangling nature of the unitary evolution in the study of measurement induced phase transitions \cite{rossini2021coherent,koh2022experimental,skinner2019measurement}.
The results presented here are contrary to that expectation as well, since indeed only in the strong measurement limit do we see effective correlations being built up in the ancilla spins causing time-crystalline behavior.

We note that our methods have elaborated how environmental engineering of the ancilla spins in a spin star model can be achieved by careful selection of the coupling Hamiltonian and measurement rates.
Our results present a novel technique of controlling the spin star model.
While inhomogenous interactions have been shown to produce complete controllability of the ancilla spins \cite{arenz2014control}, here we demonstrate that sufficient coherent control of the ancilla spins is available even in the thermodynamic limit with anisotropic interactions. Such bath engineering can be used to device technology applications such as quantum memories, which rely on coherences \cite{albert2014symmetries}. For finitely large number of spins and for finitely strong measurements, we can expect a metastable continuous time crystal whose spectral gap goes as $1/\Gamma$. We hence expect this work to benefit our foundational understanding of phase transitions and provide a concrete pathway to realistic quantum technology applications.

\begin{acknowledgments}
\paragraph{Acknowledgments.---} M.K. acknowledges support by Prime Minister's Research Fellowship (PMRF) offered by the Ministry of Education, Govt. of India. M.H. is supported by MEXT Quantum Leap Flagship Program Grants No. JPMXS0118067285 and No. JPMXS01 20319794. S.V. acknowledges support from a DST-SERB Early Career Research Award (ECR/2018/000957) and DST-QUEST grant number DST/ICPS/QuST/Theme-4/2019. S.V. thanks L. Garbe, S.R. Hassan, V.R. Krithika, T.S. Mahesh, V. Mukherjee, B. Prasanna Venkatesh, and J. Thingna for useful discussions.
\end{acknowledgments}

\newpage
\appendix
\maketitle
\onecolumngrid

\section{Appendix A: Derivation of ancilla master equation}\label{sec:EME}

In this section, we outline the derivation of Eq.~(\ref{eq:EME}) from Eq.~(\ref{eq:LME}).
Consider a bipartite open system with subsystems A and B evolving under the following master equation:
\begin{equation}\label{ME}
\frac{d\rho_{AB}}{dt} = \mathcal{K}(\rho_{AB}) +\Gamma\mathcal{D}(\rho_{AB}) = \mathcal{L}\rho_{AB}.
\end{equation}
Here $\rho_{AB}$ is the composite state, $\mathcal{K}(.)=-i[H,.]$ where $H$ is the Hamiltonian.
We consider the scenario where the dissipative superoperator $\mathcal{D}$ acts only on subsystem A and has a unique steady state.
Let $\{\rho_{k}\}$ ($\{\pi_{k}\}$) be the set of right (left) eigenvectors of the dissipator on subsystem A with eigenvalues $\{\lambda_{k}\}$.
The Zeno subspace is the steady-state manifold of the dissipator given as the kernel of dissipation superoperator, $\text{Ker} \; {\mathcal{D}} : = \rho_{0} \otimes \mathbf{B}(\mathcal{H}_{B})$.
Here $\rho_{0}$ is the steady state of the dissipator on subsystem A and $\mathbf{B}(\mathcal{H}_{B})$ is the Liouville space of subsystem B.
We are interested in evaluating the the dynamics of the system in the Zeno subspace, hence we define the projection operator onto the Zeno subspace as $\mathcal{P}(X) = \rho_{0}^{c} \otimes \text{Tr}_{c} ((\pi_{0}^{c}\otimes I^{\otimes N})X) = \rho_{0}^{c} \otimes \text{Tr}_{c}(X)$.
When the dissipation is strong, the system settles into the Zeno subspace over a fast timescale $\mathcal{O}(1/\Gamma)$ and subsequently undergoes slow dynamics in the Zeno subspace.
The separation in timescales allows us to employ generalised Schrieffer-Wolff transformation perturbatively \cite{kessler2012generalized} to decouple the fast and slow dynamics.
In the Zeno limit of strong dissipation, the effective decoupled Liouvillian describing the slow dynamics becomes,
\begin{equation}\label{eq:ELME}
\mathcal{L}_{eff} = \mathcal{P} \mathcal{K}\mathcal{P} + \frac{1}{\Gamma} \mathcal{P}\mathcal{K}\mathcal{Q}\mathcal{S}\mathcal{K}\mathcal{P} +\mathcal{O}\Big(\frac{\mathcal{K}^{3}}{\Gamma^{2}}\Big).
\end{equation}
Here $\mathcal{Q} = \mathcal{I} - \mathcal{P}$ is the orthogonal projection to the Zeno subspace and $\mathcal{S}$ is the pseudoinverse of the dissipation superoperator. 

In order to analyze the validity of Eq.~(\ref{eq:ELME}) in the thermodynamic limit, we rescale time as $\tau = \Gamma t$ and write $\Gamma = \gamma N$ where $N$ is the system size. The effective master equation in the rescaled time becomes,
\begin{equation}\label{eq:SELME}
\frac{d\rho_{AB}}{d\tau} = \mathcal{P} \frac{\mathcal{K}}{\Gamma}\mathcal{P} +  \mathcal{P}\frac{\mathcal{K}}{\Gamma}\mathcal{Q}\mathcal{S}\frac{\mathcal{K}}{\Gamma}\mathcal{P} +\mathcal{O}\Big(\frac{\mathcal{K}^{3}}{\Gamma^{3}}\Big).
\end{equation}
When the Hamiltonian term $\mathcal{K}$ scales extensively with system size $N$, $\mathcal{K}/\Gamma \sim \gamma_{0}/\gamma$ with $\gamma_{0}$ being the supremum of the parameters in $\mathcal{K}$. Thus the higher-order terms in Eq.~(\ref{eq:SELME}) and hence Eq.~(\ref{eq:ELME}) can be neglected when $\gamma_{0} \ll \gamma$.

The Hamiltonian of the entire system can be expressed in $\{\pi_{k}\}$ basis for Liouville space of subsystem A as $ H = \sum_{k} \pi_{k}^{\dagger} \otimes L_{k}$ where $L_{k} = \text{Tr}(\rho_{k} H)$.
The first term  in Eq.~(\ref{eq:ELME}) leads the unitary part of the evolution \cite{popkov18,Zanardi} with corresponding Hamiltonian given by $H_{Z} = L_{0}$.
The second term on the other hand contains the Lamb shift and dissipation terms \cite{popkov18}.
The Lamb shift Hamiltonian can be expressed as $H_{L} = \sum_{m,n>0} h_{mn} L_{m}^{\dagger}L_{n}$, with $h_{mn} = (A_{mn} - A_{nm}^{*})/2i $ where $A_{mn}=-\text{Tr}(\pi_{m}\pi_{n}^{\dagger}\rho_{0})/\lambda_{m}^{*}$.
The dissipation term has the GKLS structure with $\{L_{k}\}$ being the Lindblad operators for all non zero indices.
Thus the effective master equation for the reduced state of subsystem B becomes,
\begin{equation}
\dot{\rho}_{B} = -i[H+\frac{H_{L}}{\Gamma}, \rho_{B}] +\sum_{m,n>0} \frac{1}{\Gamma} k_{mn} (L_{n}\rho_{B}L_{m}^{\dagger} - \frac{1}{2}\{ L_{m}^{\dagger}L_{n},\rho_{B}\}).
\end{equation}
The elements of the Kossakowski matrix are $k_{mn} = A_{mn} + A_{nm}^{*} $. 
For the  spin star model with strong measurement on the central spin considered in the main text, the eigenvalues and corresponding left and right eigenvectors of the dissipator are,
\begin{subequations}
\begin{align}
    \lambda_{0} &= 0 \quad\text{ with } \rho_{0} = \ket{0}\bra{0} \text{ and }   \pi_{0} = \ket{0}\bra{0}+\ket{1}\bra{1} = I,\\
    \lambda_{1} &= -\frac{1}{2} \text{ with } \rho_{1} = \ket{0}\bra{1} \text{ and }   \pi_{1} = \ket{0}\bra{1},\\
    \lambda_{2} &= -\frac{1}{2} \text{ with } \rho_{2} = \ket{1}\bra{0} \text{ and }   \pi_{2} = \ket{1}\bra{0},\\
    \lambda_{3} &= 1 \quad \text{ with } \rho_{3} =  \ket{1}\bra{1}-\ket{0}\bra{0}\text{ and }   \pi_{3} =  \ket{1}\bra{1}.
\end{align}
\end{subequations}
 All $A_{mn}$ except $A_{11} = 1$ are zero, hence the Lamb shift term for our model vanishes and the only non-zero element in the Kossakowski matrix is $k_{11} = 4$.
 The effective Zeno Hamiltonian becomes $H_{Z} =  \text{Tr}(\ket{0_{c}}\bra{0_{c}} H) = \omega_{a} \hat{I}_{z} - \sum_{\alpha} J_{z\alpha} I_{\alpha}$.
The dissipation happens via a single channel generated by the Lindblad operator $L_{1} = \text{Tr}(\ket{0_{c}}\bra{1_{c} }H) = \sum_{\alpha \beta} (J_{x\alpha}I_{\alpha} - iJ_{y\beta}I_{\beta})/2$ with decay rate $4/\Gamma$. This reduces to Eq.~(\ref{eq:EME}) in the main text, justifying the master equation considered in the main text in the limit of strong measurement of the central spin.

\section{Appendix B: Meanfield analysis of the spin star system}\label{sec:meanfield}
In this section, we derive the meanfield dynamics of the spin star variables for pedagogical understanding.
Following the master equation of the spin star system given by Eq.~(\ref{eq:LME}), expectation values of the central spin operators $\langle S_{\alpha} \rangle$ and the ancillary spin operators $\langle I_{\alpha} \rangle$ results in a system of coupled differential equations as given below 
\begin{subequations}
\begin{align}
\langle \Dot{I}_x \rangle&=-\omega_a \langle I_y \rangle+J_{yy}\langle S_y \rangle \langle I_z \rangle-J_{zz} \langle S_z \rangle \langle I_y \rangle,\label{eq:mfa} \\
\langle \Dot{I}_y \rangle&=\omega_a\langle I_x \rangle+J_{zz}\langle S_z \rangle\langle I_x \rangle-J_{xx}\langle S_x \rangle \langle I_z \rangle- J_{zx} \langle S_z \rangle \langle I_z \rangle,\label{eq:mfb} \\
\langle \Dot{I}_z \rangle&=J_{xx}\langle S_x \rangle \langle I_y \rangle- J_{yy}\langle S_y \rangle \langle I_x \rangle +J_{zx} \langle S_z \rangle \langle I_y \rangle,\label{eq:mfc} \\
\langle \Dot{S}_x \rangle&=-\omega_c \langle S_y \rangle+J_{yy}\langle S_z \rangle\langle I_{y} \rangle +J_{zx} \langle S_y \rangle \langle I_{x} \rangle - J_{zz} \langle S_y \rangle \langle I_{z} \rangle - \frac{\Gamma}{2} \langle S_x \rangle,\label{eq:mfd}  \\
\langle \Dot{S}_y \rangle&=-\omega_c \langle S_x \rangle- J_{xx}\langle S_z \rangle \langle I_{x} \rangle+J_{zx} \langle S_x \rangle \langle I_{x} \rangle+ J_{zz} \langle S_x \rangle \langle I_{z} \rangle - \frac{\Gamma}{2} \langle S_y \rangle,\label{eq:mfe} \\
\langle \Dot{S}_z \rangle&=J_{xx}\langle S_y \rangle \langle I_{x} \rangle -J_{yy} \langle S_x \rangle \langle I_{y} \rangle - \Gamma (\langle S_z \rangle+\frac{1}{2})\label{eq:mff}.
\end{align}
\end{subequations}

Since the central spin is subjected to strong measurements $\Gamma \rightarrow \infty$, there is a timescale separation between the dynamics of the central and the ancilla spins.
Therefore, we can use adiabatic elimination to solve these coupled non-linear equations \cite{haken1977synergetics}.
The central spin settle down to the fixed point provided by $\langle \Dot{S}_x\rangle =\langle \Dot{S}_y\rangle =\langle \Dot{S}_z\rangle =0$ in short time span of $\mathcal{O}(\frac{1}{\Gamma})$. Corresponding fixed point of the central spin is given by
\begin{subequations}
\begin{align}
\langle S_x\rangle &= -\frac{ \Gamma (J_{yy}\Gamma\langle I_{y}\rangle  - 2 J_{xx}J_{zx} \braket{I_{x}}^2  + 2 J_{zz} \langle I_{x} \rangle \langle I_{z}\rangle +2\omega_c J_{xx} \langle I_{x}\rangle)}{\Gamma^3 +2 J_{xx}^2 \Gamma \braket{I_{x}}^2+ 8 J_{xx} J_{yy} \langle I_x \rangle \langle I_y \rangle  (- J_{zx} \langle I_x \rangle  + \omega_c) + \Gamma( 2 J_{yy}^2\langle I_y \rangle^2 + 4 J_{zz}^2 \langle I_y \rangle^2  -4 (- J_{zx} \langle I_x \rangle  + \omega_c)^2)},\nonumber  \\
\langle S_y \rangle &= \frac{ \Gamma (\Gamma J_{xx}\langle I_{x} \rangle - 2 J_{yy} \langle I_{y}\rangle(J_{zx}\langle I_{x} \rangle+J_{zz}\langle I_{z} \rangle-\omega_c))}{\Gamma^3 +2 J_{xx}^2 \Gamma \braket{I_{x}}^2+ 8 J_{xx} J_{yy} \langle I_x \rangle \langle I_y \rangle  (- J_{zx} \langle I_x \rangle  + \omega_c) + \Gamma( 2 J_{yy}^2\langle I_y \rangle^2 + 4 J_{zz}^2 \langle I_y \rangle^2  -4 (- J_{zx} \langle I_x \rangle  + \omega_c)^2)},\nonumber \\
\langle S_z \rangle &= -\frac{\Gamma (\Gamma^2 +4 J_{zz}^2 \braket{I_z}^2  -  4 (- J_{zx} \braket{I_x} +\omega_c^2))}{2 (\Gamma^3+2 J_{xx}^2 \Gamma \braket{I_{x}}^2+8J_{xx}J_{yy}\braket{I_x} \braket{I_y}(\omega_c-J_{zx}\braket{I_x})+\Gamma(2J_{yy}^2 \braket{I_y}^2+4J_{zz}^2 \langle I_z \rangle^2 -4(-J_{zx}\braket{I_x}\omega_c)^2))\nonumber },
\end{align}
\end{subequations}
which relates the central spin variables $\langle S_{\alpha}\rangle$ with the ancilla spin variables $\langle I_{\alpha}\rangle$. In the large $\Gamma$ limit, the above fixed point simplifies to
\begin{equation}
    \{\langle S_x\rangle,\langle S_y\rangle,\langle S_z\rangle\}\approx \Big\{-\frac{J_{yy}}{\Gamma}\langle I_{y}\rangle,\frac{J_{xx}}{\Gamma}\langle I_{x}\rangle,-\frac{1}{2}\Big\}.
\end{equation}
After substituting $\Gamma=\gamma I$, above equation can be rewritten in terms of the rescaled spin variables $m_{\alpha}=\langle I_{\alpha} \rangle/I$ as given below
\begin{equation}
    \{\langle S_x\rangle,\langle S_y\rangle,\langle S_z\rangle\}\approx  \Big\{-\frac{J_{yy}}{\gamma}m_y,  \frac{J_{xx}}{\gamma}m_x,  -\frac{1}{2}\Big\}.\label{eq:fixed}
\end{equation}
In the Zeno limit, where the central spin is strongly coupled to the bath $\gamma \rightarrow \infty$, the central spin settles down to the ground state as pointed out by the fixed point $\{\langle S_x\rangle,\langle S_y\rangle,\langle S_z\rangle\}=\{0,0,-1/2\}$ given by Eq.~(\ref{eq:fixed}). Substituting above equation in Eqs.~(\ref{eq:mfa}-\ref{eq:mfc}), the meanfield equations for the rescaled ancillary spin operators are given as follow,
\begin{subequations}
\begin{align}
    \Dot{m}_x &=\frac{J_{xx}J_{yy}}{\gamma} m_x m_z + \Big(\frac{J_{zz}}{2}-\omega_a\Big)m_y,\\
    \Dot{m}_y &= \frac{J_{xx}J_{yy}}{\gamma} m_y m_z +\frac{J_{zx}}{2} m_z-\Big(\frac{J_{zz}}{2}-\omega_a\Big) m_x,\\
    \Dot{m}_z &= -\frac{J_{xx}J_{yy}}{\gamma} (m_x^2+ m_y^2)-\frac{J_{zx}}{2} m_y.
\end{align}
\end{subequations}
For the choice of the coupling $\mathbf{J}$ matrix described in section on the thermodynamic limit in the main text, namely ($J_{xx} = J_{yy}, J_{zz} = 2\omega_{a} \text{ and } J_{zx} = -2\Omega$), above meanfield equations can be rewritten as
\begin{subequations}
\begin{align}
    \Dot{m}_x &=\kappa m_x m_z,\label{eq:ztc_mfa} \\
    \Dot{m}_y &= \kappa m_y m_z -\Omega m_z,\label{eq:ztc_mfb}\\
    \Dot{m}_z &= -\kappa (m_x^2+ m_y^2)+\Omega m_y,\label{eq:ztc_mfc}
\end{align}
\end{subequations}
where $\kappa=J_{xx}^{2}/\gamma$. These equations describe the meanfield dynamics of the boundary time crystals discussed in \cite{iemini2018boundary}. This is consistent with the reduced master equation of the ancillary spins given by Eq.~(\ref{eq:ZTC}) in the main text which describe the time evolution of a boundary time crystal.
The operator $I^2=I_x^2+I_y^2+I_z^2$ commutes with all other spin operators and hence is a strong symmetry of the system.
Therefore total spin $I^2$ is a conserved quantity of the system and is also conserved by Eqs.~(\ref{eq:ztc_mfa}-\ref{eq:ztc_mfc}). Fixed points for the above  given set of equations along with $m_x^2+m_y^2+m_z^2=1$ are given by
\begin{subequations}
\begin{align}
    \{m_x^{(1)},m_y^{(1)},m_z^{(1)}\}=\Big\{\pm \sqrt{1-\left(\frac{\kappa}{\Omega}\right)^2},-\frac{\kappa}{\Omega},0\Big\}, \\
    \{m_x^{(2)},m_y^{(2)},m_z^{(2)}\}=\Big\{0,-\frac{\Omega}{\kappa},\pm \sqrt{1-\left(\frac{\Omega}{\kappa}\right)^2}\Big\}.
\end{align}
\end{subequations}
Stability of the fixed point is given by the eigenvalues of the Jacobian matrix which gives the linearisation around the given fixed point.
Eigenvalues of the Jacobian matrix for the given fixed points $\{m_x^{(1)},m_y^{(1)},m_z^{(1)}\}$ and $\{m_x^{(2)},m_y^{(2)},m_z^{(2)}\}$ are $\{\lambda^{(1)}\}=\{0,\pm\sqrt{\kappa^2-\Omega^2}\}$ and $\{\lambda^{(2)}\}=\{0,\sqrt{\kappa^2-\Omega^2},\sqrt{\kappa^2-\Omega^2}\}$ respectively.
We can observe that for $\Omega/\kappa < 1$, $\{m_x^{(1)},m_y^{(1)},m_z^{(1)}\}$ is a physical fixed point of the system and corresponding Jacobian eigenvalues are real which makes it a saddle fixed point. Hence for $\Omega/\kappa < 1$, system does not oscillate and settle down to $\{m_x^{(1)},m_y^{(1)},m_z^{(1)}\}$ fixed point.
On the other hand, for $\Omega/\kappa >1$, $\{m_x^{(1)},m_y^{(1)},m_z^{(1)}\}$ becomes unphysical and $\{m_x^{(2)},m_y^{(2)},m_z^{(2)}\}$ is new physical fixed point of the system. Corresponding Jacobian eigenvalues for $\Omega/\kappa >1$ are either zero or purely imaginary which corresponds to centers. This give rise to stable oscillations of rescaled spin operators and hence breaks the continuous time translational symmetry of the system in thermodynamic limit for $\Omega/\kappa >1$.

\end{document}